\documentclass[aps,prc,twocolumn,showpacs,superscriptaddress,floatfix,10pt]{revtex4-1}
\usepackage{bm}
\usepackage{dcolumn}
\usepackage{amsthm}
\usepackage{amssymb}
\usepackage{amsmath}
\usepackage{graphicx}
\usepackage{bm}
\usepackage{xcolor}
\usepackage{mathptmx} 
\usepackage[colorlinks,allcolors=blue]{hyperref}
\usepackage{color}

\begin{document}

\title{Fusion and quasifission dynamics in the reactions $^{48}$Ca+$^{249}$Bk
       and $^{50}$Ti+$^{249}$Bk using TDHF}

\author{A.S. Umar}\email{umar@compsci.cas.vanderbilt.edu}
\author{V.E. Oberacker}\email{volker.e.oberacker@vanderbilt.edu}
\affiliation{Department of Physics and Astronomy, Vanderbilt University, Nashville, Tennessee 37235, USA}
\author{C. Simenel}\email{cedric.simenel@anu.edu.au}
\affiliation{Department of Nuclear Physics, Research School of Physics and Engineering, The Australian National University, Canberra ACT  2601, Australia}
\date{\today}


\begin{abstract}
\begin{description}
\item[Background]
Synthesis of superheavy elements (SHE) with fusion-evaporation reactions is strongly hindered by the 
quasifission (QF) mechanism which prevents the formation of an equilibrated compound nucleus and which depends on the structure of the reactants. 
New SHE have been recently produced with doubly-magic $^{48}$Ca beams. 
However, SHE synthesis experiments with single-magic $^{50}$Ti beams have so far been unsuccessful.
\item[Purpose]
In connection with experimental searches for $Z=117,119$ superheavy elements, we perform a theoretical study of
fusion and quasifission mechanisms in $^{48}$Ca,$^{50}$Ti+$^{249}$Bk
 reactions in order to investigate possible differences in reaction mechanisms induced by these two projectiles. 
 \item[Methods]
The collision dynamics and the outcome of the reactions are studied using unrestricted time-dependent Hartree-Fock (TDHF) calculations 
as well as the density-constrained TDHF method to extract the nucleus-nucleus potentials and the excitation energy in each fragment. 
\item[Results] Nucleus-nucleus potentials, nuclear contact times, masses and charges of the  fragments, as well as their kinetic and excitation energies strongly depend on the orientation of the prolate $^{249}$Bk nucleus.
Long contact times associated with fusion are observed in collisions of both projectiles with the side of  the $^{249}$Bk nucleus, but not on collisions with its tip. 
The energy and impact parameter dependences  of the fragment properties, as well as their mass-angle and mass-total kinetic energy correlations  are  investigated.
\item[Conclusions]
Entrance channel reaction dynamics are similar with both $^{48}$Ca and $^{50}$Ti projectiles. Both are expected to lead to the formation of a compound nucleus by fusion if they have enough energy to get in contact with the side of the $^{249}$Bk target.
\end{description}
\end{abstract}
\maketitle


\section{Introduction}

The synthesis of superheavy elements is one of the most fascinating and challenging tasks
in low-energy heavy-ion physics. Nuclear mean-field theories predict a
superheavy island of stability as a result of new proton and neutron shell closures.
Most recent theoretical calculations yield a magic neutron number $N=184$, but there is no
consensus yet about the corresponding magic proton number, with
predictions~\cite{bender1999, bender2001, nazarewicz2002, cwiok2005, pei2005, pei2009a}
ranging from $Z=114-126$.
Experimentally, two approaches have been used for the synthesis of these elements.
The first method uses targets containing doubly-magic spherical nuclei such as
$^{208}$Pb (or alternatively $^{209}$Bi). By bombarding these targets with heavy-ion beams ranging from chromium
to zinc, researchers at the GSI Helmholtz Center in Germany and at Riken were able to produce
several isotopes of elements $Z=107-112$. The beam energy was kept low to
minimize the excitation energy (`cold fusion')~\cite{hofmann2000,hofmann2002,munzenberg2015,morita2015}.
The second approach, pioneered at JINR in Russia, uses actinide targets instead.
In contrast to the spherical $^{208}$Pb target nuclei used at GSI, all
of the actinide target nuclei exhibit quadrupole deformed ground states.
Target materials ranging from $^{238}$U to $^{249}$Cf were irradiated with
a $^{48}$Ca beam. Despite the fact that the excitation energy is found to be
substantially higher in these experiments (`hot-fusion') researchers at JINR were able to create
isotopes of elements $Z=113-118$~\cite{oganessian2007,oganessian2010,oganessian2013,roberto2015,oganessian2015},
with lifetimes of milliseconds up to a minute. Recently, hot-fusion
experiments were also carried out at GSI, LBNL, and
RIKEN~\cite{hofmann2007,stavsetra2009,dullmann2010,hofmann2012,rudolph2013,khuyagbaatar2014,morita2015}
which confirmed the discovery of elements $Z=112-117$.

However, attempts to synthesize even heavier elements such as $Z=119,120$ with beams
of $^{50}$Ti and $^{54}$Cr instead of $^{48}$Ca have so far not been successful.
The experimental community is asking for theoretical guidance as to
why $^{48}$Ca beams seem to be so crucial in forming superheavy elements.
For example, the reaction $^{48}$Ca +$^{249}$Bk produces superheavy
element $117$ with cross-sections of $2-3$ picobarns. By contrast,
an upper cross section limit of only
$50$~fb was reported for the production of isotopes of element $119$
in the reaction $^{50}$Ti+$^{249}$Bk at GSI-TASCA~\cite{dullmann2013}.

Experimentally it is found that capture reactions involving actinide target
nuclei result either in fusion or in quasifission. Fusion produces a compound
nucleus in statistical equilibrium, while quasifission leads to a
reseparation of the fragments after partial mass equilibration without
formation of an equilibrated compound nucleus  \cite{bock1982}. 
Furthermore, if the nucleus does not quasifission and evolves to a compound system,
it can still undergo statistical fission due to its excitation.
The evaporation residue cross-section
is dramatically reduced due to the quasifission (QF) and fusion-fission (FF)
processes.

Quasifission occurs at a much shorter time-scale than fusion-fission \cite{toke1985,shen1987,rietz2011}.
Consequently, quasifission is the primary reaction
mechanism that limits the formation of superheavy nuclei \cite{sahm1984,gaggeler1984,schmidt1991}. 
This motivated intensive experimental studies
\cite{toke1985,shen1987,hinde1992,hinde1995,hinde1996,itkis2004,knyazheva2007,hinde2008,nishio2008,kozulin2014,rietz2011,itkis2011,lin2012,nishio2012,simenel2012b,rietz2013,williams2013,kozulin2014,wakhle2014,hammerton2015,prasad2015,prasad2016}. 
These studies have shown a strong impact of the entrance channel characteristics, including deformation \cite{hinde1995,hinde1996,knyazheva2007,hinde2008,nishio2008} and shell structure \cite{simenel2012b} of the reactants. 
The later stages of the dynamics are also impacted by the fissility of the total system \cite{lin2012,rietz2013}, its neutron richness \cite{hammerton2015},  and by shell effects in the exit channel \cite{toke1985,shen1987,itkis2004,nishio2008,morjean2008,fregeau2012,kozulin2010,kozulin2014,wakhle2014}.

Most dynamical models~\cite{fazio2005,adamian2003,adamian2009,nasirov2009a,feng2009}
argue that for heavy systems a dinuclear
complex is formed initially and  the barrier structure and the excitation energy of this precompound
system will determine its survival to breaking up via  quasi-fission.
The challenge for nuclear theory is to describe the entrance channel dynamics 
leading either to fusion or to quasifission and which accounts for the complex interplay between dynamics and structure.
Microscopic dynamical theories are natural candidates to describe such reactions. 
Here, we simulate heavy-ion collisions in the framework of the  time-dependent Hartree-Fock (TDHF) theory which provides a fully microscopic mean-field approach to nuclear dynamics. 

In this paper we will concentrate on the theoretical analysis of the
$^{48}$Ca+$^{249}$Bk experiments~\cite{oganessian2010,oganessian2013,khuyagbaatar2014}
in which element $Z=117$ was produced.
This system will be compared to the $^{50}$Ti+$^{249}$Bk reaction which appears to have
a very low cross section limit for synthesizing element $Z=119$.
Our goal is to investigate potential different mechanisms between these two reactions by calculating dynamical observables such as nuclear contact times,
mass and charge transfer, excitation energies, and heavy-ion potentials.

A brief introduction to the theoretical framework is provided in section~\ref{sec:TDHF}, followed by a presentation and discussion of the results in section~\ref{sec:results}.
Conclusions are drawn in section~\ref{sec:conclusions}.

\section{Formalism: TDHF and DC-TDHF} \label{sec:TDHF}

The Time-Dependent Hartree-Fock (TDHF) theory allows us to study a large variety of phenomena
observed in low energy nuclear physics~\cite{negele1982,simenel2012}.
In particular, TDHF provides a dynamic quantum many-body description of large amplitude collective
motion including collective surface vibrations and giant
resonances~\cite{umar2005a,maruhn2005,nakatsukasa2005,simenel2003,
reinhard2006,reinhard2007,simenel2009,fracasso2012,pardi2013,pardi2014,suckling2010,stetcu2011,avez2013,scamps2014a},
nuclear reactions in the
vicinity of the Coulomb barrier, such as fusion~\cite{bonche1978,flocard1978,simenel2001,
washiyama2008,umar2010a,guo2012,keser2012,simenel2013a,oberacker2012,oberacker2010,umar2012a,simenel2013b,umar2014a,jiang2014},
deep-inelastic reactions and transfer~\cite{koonin1977,simenel2010,simenel2011,umar2008a,
sekizawa2013,scamps2013a,sekizawa2015,bourgin2016},
and dynamics of (quasi)fission fragments~\cite{wakhle2014,oberacker2014,simenel2014a,
umar2015a,umar2015c,scamps2015a,goddard2015,bulgac2016,sekizawa2016}.

The TDHF equations for the single-particle wave functions
\begin{equation}
h(\{\phi_{\mu}\}) \ \phi_{\lambda} (r,t) = i \hbar \frac{\partial}{\partial t} \phi_{\lambda} (r,t)
            \ \ \ \ (\lambda = 1,...,A) \ ,
\label{eq:TDHF}
\end{equation}
can be derived from a variational principle. The main approximation in TDHF is
that the many-body wave function $\Phi(t)$  is assumed to be a single time-dependent
Slater determinant at all times. It describes the time-evolution of the single-particle
wave functions in a mean-field corresponding to the dominant reaction channel.
During the past decade it has become numerically feasible to perform TDHF calculations on a
3D Cartesian grid without any symmetry restrictions
and with much more accurate numerical methods~\cite{reinhard1988,umar1989,bottcher1989,nakatsukasa2005,umar2006c,sekizawa2013,maruhn2014,simenel2012}.
Furthermore, the quality of effective interactions has been substantially
improved~\cite{chabanat1998a,guichon2006,kluepfel2009,kortelainen2010}.

Recently, we have developed a new dynamic microscopic approach,
the density-constrained time-dependent Hartree-Fock (DC-TDHF)
method~\cite{umar2006b}, to calculate nucleus-nucleus potentials $V(R)$, mass parameters $M(R)$, and
precompound excitation energies $E^{*}(R)$~\cite{umar2009a}, directly from microscopic TDHF dynamics.
The basic idea of this approach is the following:
At certain times $t$ or, equivalently, at certain internuclear distances
$R(t)$ the instantaneous TDHF density is used to perform a static energy minimization
while constraining the proton and neutron densities to be equal to the instantaneous
TDHF densities. This can be accomplished by solving the density-constrained density-functional
problem
\begin{widetext}
\begin{equation}
E_{DC}(t)={\min_{\rho_n,\rho_p}} \left\{ E[\rho_n,\rho_p]+\int d\mathbf{r}\, v_n(\mathbf{r})\left[\rho_n(\mathbf{r})-
\rho_{n}^{tdhf}(\mathbf{r},t)\right]+\int d\mathbf{r}\, v_p(\mathbf{r})\left[\rho_p(\mathbf{r})-
\rho_{p}^{tdhf}(\mathbf{r},t)\right]\right\}\;,
\label{edc}
\end{equation}
\end{widetext}
where $E[\rho_n,\rho_p]$ is the TDHF density-functional (calculated with Skyrme interactions).
The quantities $v_{n,p}(\mathbf{r})$ are Lagrange multipliers which represent external
fields that constrain the densities during the minimization procedure.
This means we
allow the single-particle wave functions to rearrange themselves in such a way
that the total energy is minimized, subject to the TDHF density constraint.
In a typical DC-TDHF run, we utilize a few
thousand time steps, and the density constraint is applied every $10-20$ time steps.
We refer to the minimized energy as the ``density constrained energy''
$E_{\mathrm{DC}}(R(t))$.
The ion-ion interaction potential $V(R)$ is obtained by
subtracting the constant binding energies
$E_{\mathrm{A_{1}}}$ and $E_{\mathrm{A_{2}}}$ of the two individual nuclei
\begin{equation}
V(R)=E_{\mathrm{DC}}(R)-E_{\mathrm{A_{1}}}-E_{\mathrm{A_{2}}}\ .
\label{eq:vr}
\end{equation}
The calculated ion-ion interaction barriers contain all of the dynamical changes in the nuclear
density during the TDHF time-evolution in a self-consistent manner.

In addition to the ion-ion potential it is also possible to obtain coordinate
dependent mass parameters. One can compute the ``effective mass'' $M(R)$~\cite{umar2009b}
using the conservation of energy in a central collision
\begin{equation}
M(R)=\frac{2[E_{\mathrm{c.m.}}-V(R)]}{\dot{R}^{2}}\;,
\label{eq:mr}
\end{equation}
where the collective velocity $\dot{R}$ is directly obtained from the TDHF time evolution and the potential
$V(R)$ from the density constraint calculations.
At large distance $R$, the mass $M(R)$ is equal to the
reduced mass $\mu$ of the system. At smaller distances, when the nuclei overlap, the
mass parameter generally increases. This microscopic approach also applies to reactions
involving deformed nuclei when calculations are done in an unrestricted three-dimensional box where the nuclei can be given arbitrary orientations with respect to the collision axis~\cite{umar2006d,umar2008b,umar2006a}.

Using the density constrained energy defined above, we can compute the excitation energy
of the system at internuclear distance $R(t)$ as follows
\begin{equation}
E^{*}(R(t)) = E_{TDHF} - E_{\mathrm{DC}}(R(t)) - E_{kin}(R(t)) \;.
\end{equation}
where $E_{TDHF}$ is the conserved TDHF energy. The last term denotes the
collective kinetic energy
\begin{equation}
E_{kin} \approx \frac{m}{2}\sum_q\int d^{3}r\; \mathbf{j}_q(\mathbf{r},t)^2/\rho_q(\mathbf{r},t)\;,
\end{equation}
where $\mathbf{j}(\mathbf{r},t)$ is the local current density from TDHF and $m$ the nucleon mass. 
The index $q$ denotes the isospin index for neutrons and protons ($q=n,p$). 


\section{Numerical results}\label{sec:results}

\subsection{Unrestricted TDHF calculations: fusion and quasifission}

In this paper, we focus on fusion and quasifission in the reactions
$^{48}$Ca+$^{249}$Bk and $^{50}$Ti+$^{249}$Bk. In our TDHF calculations
we use the Skyrme SLy4 and SLy4d energy density functionals~\cite{chabanat1998a,kim1997}
including all of the relevant time-odd terms in the mean-field Hamiltonian. 
Both interactions are constructed using the same fitting procedure, apart for one-body center of mass corrections, included in SLy4, which are small in heavy systems such as those studied here. 
The $^{48}$Ca+$^{249}$Bk calculations were done with SLy4d while the calculations
for $^{50}$Ti+$^{249}$Bk used the SLy4 parametrization. The reason for switching to SLy4
was due to the availability of the pairing force parameters for this force.
To describe these reactions with a high degree of accuracy, the shapes of the individual
nuclei must be correctly reproduced by the mean-field theory. In some cases, it is necessary to
include BCS pairing which increases the number of single-particle levels that must
be taken into account by about 50 percent.
Static Hartree-Fock (HF) calculations without pairing predict a spherical
density distribution for $^{48}$Ca while $^{249}$Bk shows prolate quadrupole
and hexadecapole deformation, in agreement with experimental data.
However, static HF calculations without pairing predict a prolate
quadrupole deformation for $^{50}$Ti due to partial filling of $\pi f_{7/2}$ with occupation numbers 0 or 1, thus breaking spherical symmetry. When BCS pairing is added, these occupation number are lower than 1 and distributed around the Fermi surface, restoring a spherical density in $^{50}$Ti.
Therefore, we include BCS pairing (using fixed partial occupations) in the
TDHF runs for $^{50}$Ti+$^{249}$Bk while pairing has been left out for the
system $^{48}$Ca+$^{249}$Bk to speed up the calculations.

Numerically, we proceed as follows: First we generate very well-converged static
HF wave functions for the two nuclei on the 3D grid.
The initial separation of the two nuclei is $30$~fm. In the second
step, we apply a boost operator to the single-particle wave functions. The time-propagation
is carried out using a Taylor series expansion (up to orders $10-12$) of the
unitary mean-field propagator, with a time step $\Delta t = 0.4$~fm/c.
For reactions leading to superheavy dinuclear systems, the TDHF calculations
require very long CPU times: a single TDHF run at fixed $E_\mathrm{c.m.}$ energy
and fixed impact parameter $b$ takes about 1-2 weeks of CPU time
on a 16-processor LINUX workstation. A total CPU time of about 6 months was required
for all of the calculations presented in this paper.
\begin{figure}[!htb]
\includegraphics*[width=8.6cm]{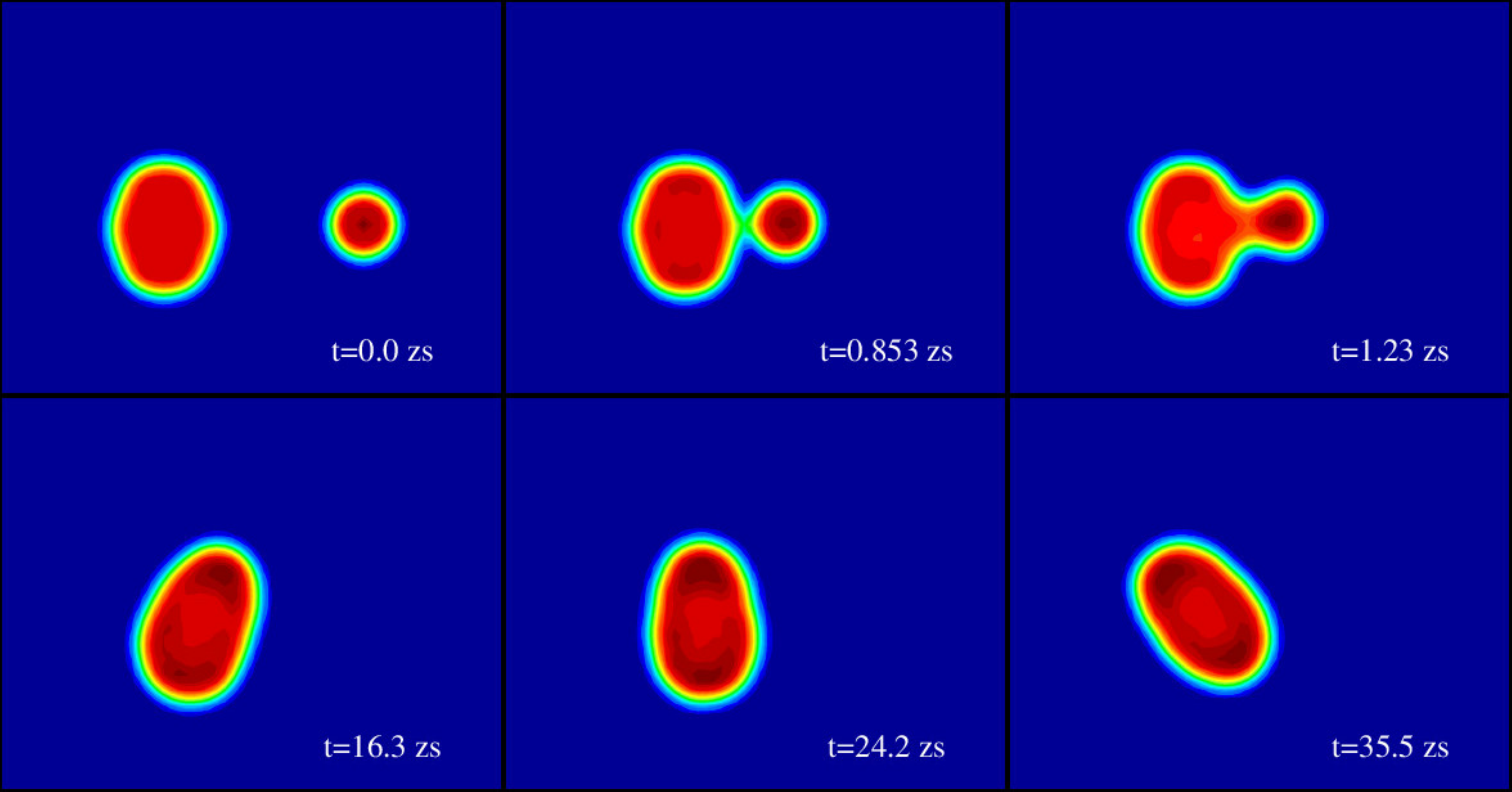}
\caption{\protect(Color online) Fusion in the reaction $^{50}$Ti+$^{249}$Bk
		at $E_{\mathrm{c.m.}}=233$~MeV with impact parameter $b=0.5$~fm.
		Shown is a contour plot of the time evolution of the mass density. Time
		increases from left to right and top to bottom.}
\label{fig:fig1}
\end{figure}

Let us first consider the reaction $^{50}$Ti+$^{249}$Bk at $E_{\mathrm{c.m.}}=233$~MeV,
which is the energy used in the GSI-TASCA experiment~\cite{dullmann2013}.
The numerical calculations were carried out on a 3D Cartesian grid which
spans $(66 \times 52 \times 30)$~fm. In Fig.~\ref{fig:fig1} we show contour plots
of the mass density in the $x-z$ plane  as a function of time.
In this case, the initial orientation of the $^{249}$Bk
nucleus has been chosen such that the $^{50}$Ti projectile collides with the
``side'' of the deformed target nucleus. We observe that
at an impact parameter $b=0.5$~fm TDHF theory predicts
fusion. Our conceptual definition of fusion is an event with
large contact time exceeding $25-35$~zs, and in addition we require a mononuclear shape
without any neck formation.

\begin{figure}[!htb]
\includegraphics*[width=8.6cm]{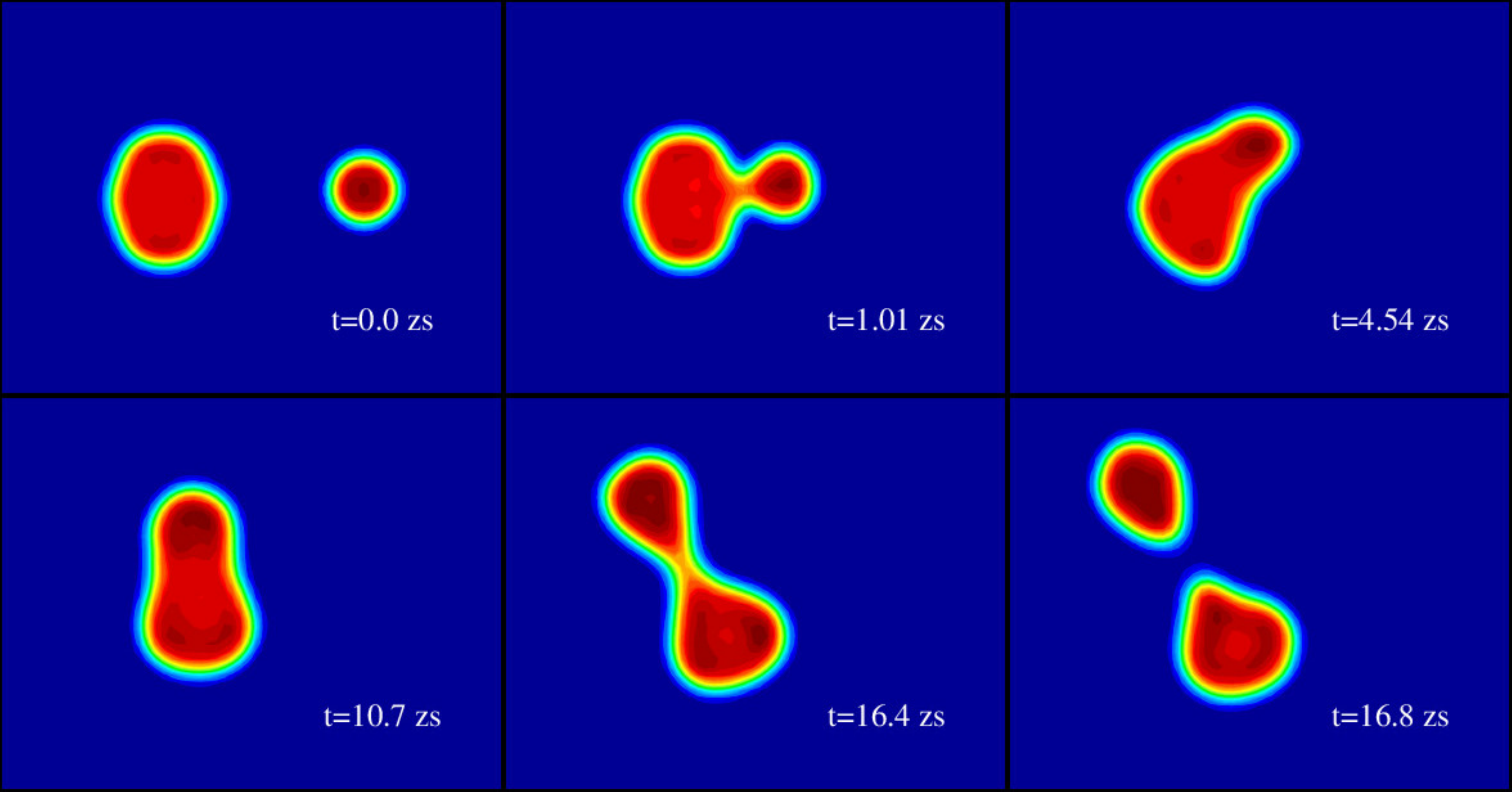}
\caption{\protect(Color online) Quasifission in the reaction $^{50}$Ti+$^{249}$Bk
		at $E_{\mathrm{c.m.}}=233$~MeV with impact parameter $b=1.0$~fm.
		Shown is a contour plot of the time evolution of the mass density. Time
		increases from left to right and top to bottom.}
\label{fig:fig2}
\end{figure}

By contrast, at an impact parameter $b=1.0$~fm TDHF theory predicts
quasifission, see Fig.~\ref{fig:fig2}.
As the nuclei approach each other, a neck forms between the
two fragments which grows in size as the system begins to rotate.
Due to the Coulomb repulsion and centrifugal forces,
the dinuclear system elongates and forms a very long neck which eventually
ruptures leading to two separated fragments.
In this case, the contact time is found
to be $16$~zs.


\subsection{Nucleus-nucleus potentials (DC-TDHF)}

\begin{figure}[!htb]
\includegraphics*[width=8.6cm]{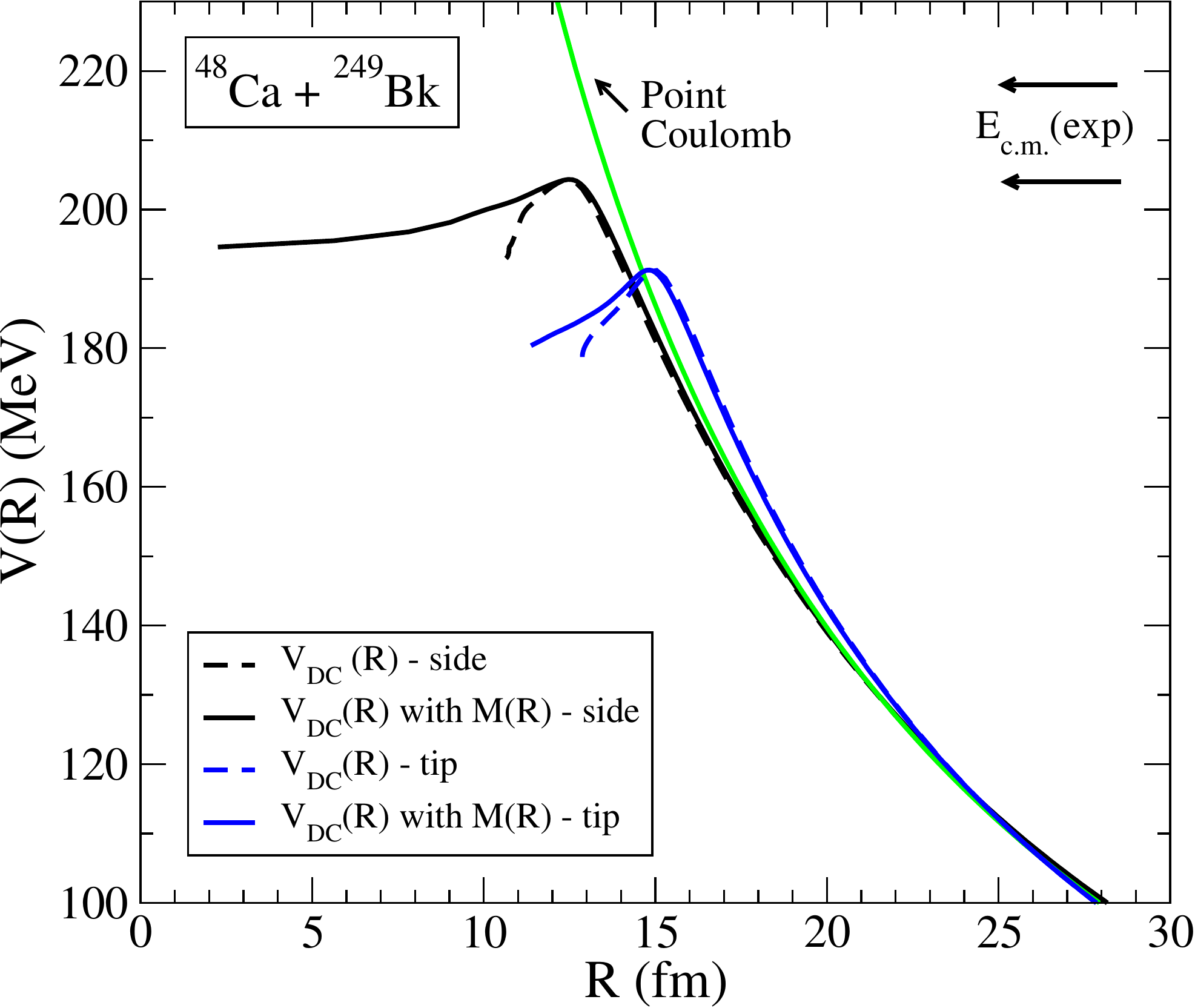}
\caption{\protect(Color online) Nucleus-nucleus potential, $V(R)$, for the $^{48}$Ca+$^{249}$Bk system
obtained from DC-TDHF calculation for selected orientation angles of the $^{249}$Bk nucleus.
Also shown is the range of the experimental c.m. energies.}
\label{fig:fig3}
\end{figure}
In Fig.~\ref{fig:fig3} we plot the microscopic DC-TDHF nucleus-nucleus potential barriers
for the $^{48}$Ca+$^{249}$Bk system. The dashed lines correspond to potentials calculated
with constant reduced mass, while the solid lines include the influence of the coordinate-dependent
``effective mass'' $M(R)$. We observe that the coordinate-dependent mass changes only
the interior region of the potential barriers.
The barriers are depicted for two extreme orientations
of the $^{249}$Bk nucleus (tip and side). As expected, the tip orientation of $^{249}$Bk
results in a significantly lower barrier, $E_B$(tip)$=191.22$~MeV located at internuclear
distance $R_B$(tip)$=15.04$~fm, as compared to the
side orientation, $E_B$(side)$=204.36$~MeV with $R_B$(side)$=12.47$~fm. (For comparison,
in the phenomenological Bass model for two spherical nuclei one obtains a barrier height
$E_B$(Bass)$=203.1$~MeV located at an internuclear distance $R_B$(Bass)$=12.8$~fm.)
Also shown is the range of experimental energies at which this reaction
has been studied, $E_\mathrm{c.m.}=204-218$~MeV in the Dubna experiment~\cite{oganessian2013} and
$E_\mathrm{c.m.}=211-218$~MeV in the GSI-TASCA experiment~\cite{khuyagbaatar2014}.
We conclude that the highest experimental energy $E_\mathrm{c.m.}=218$~MeV is
above both barriers but the lowest experimental energy $E_\mathrm{c.m.}=204$~MeV
is slightly below the barrier for the side orientation of $^{249}$Bk.

\begin{figure}[!htb]
\includegraphics*[width=8.6cm]{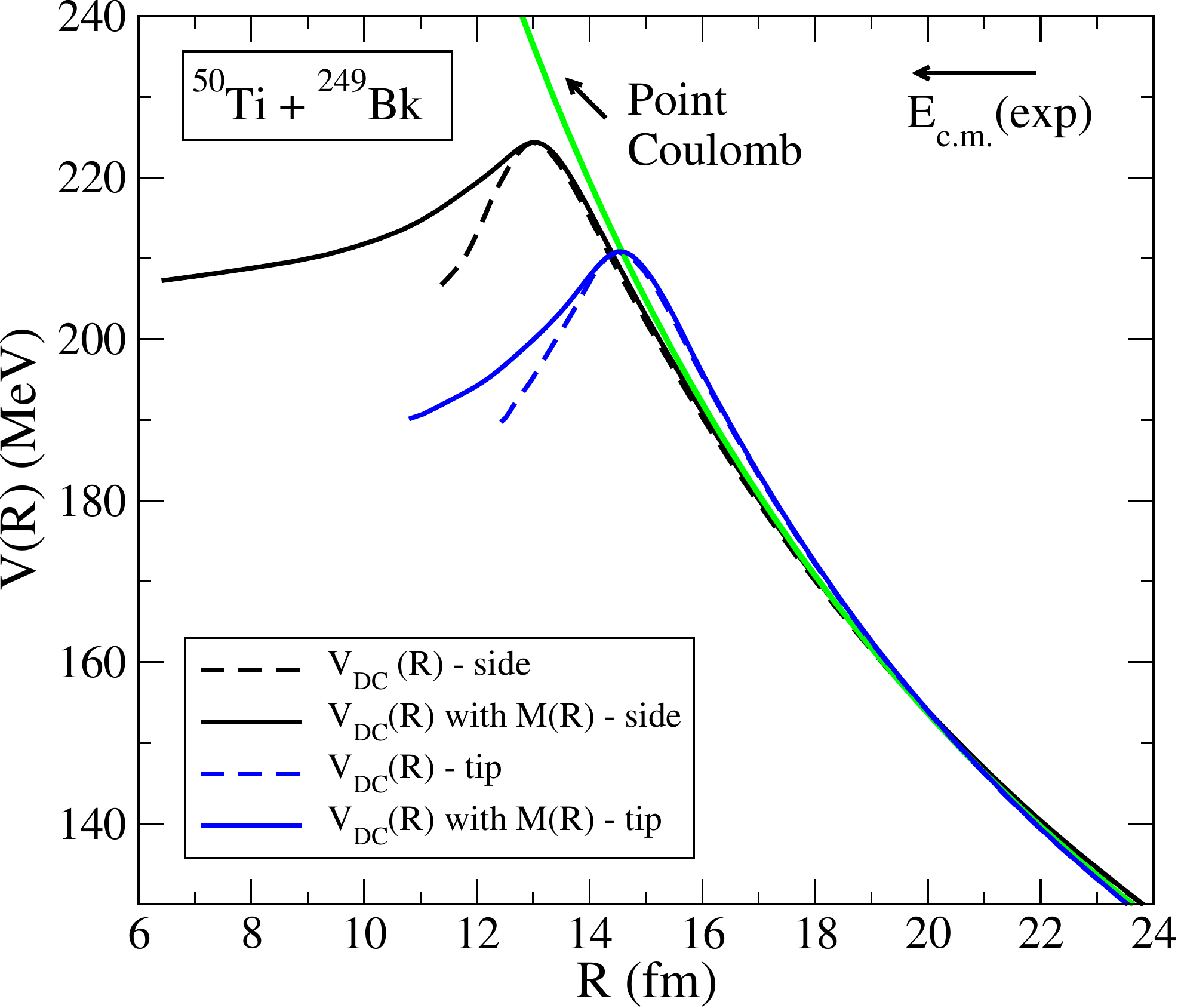}
\caption{\protect(Color online) Nucleus-nucleus potential, $V(R)$, for the $^{50}$Ti+$^{249}$Bk system
obtained from DC-TDHF calculation for selected orientation angles of the $^{249}$Bk nucleus.
Also shown is the energy $E_\mathrm{c.m.}=233.2$~MeV used in the GSI-TASCA experiment.}
\label{fig:fig4}
\end{figure}
In Fig.~\ref{fig:fig4} we plot the corresponding potential barriers
for the $^{50}$Ti+$^{249}$Bk system. Again, the tip orientation of $^{249}$Bk
results in a significantly lower barrier, $E_B$(tip)$=211.2$~MeV located at internuclear
distance $R_B$(tip)$=14.48$~fm, as compared to the
side orientation, $E_B$(side)$=224.6$~MeV with $R_B$(side)$=12.96$~fm. (For comparison,
in the phenomenological Bass model for two spherical nuclei one obtains a barrier height
$E_B$(Bass)$=223.7$~MeV located at an internuclear distance $R_B$(Bass)$=12.8$~fm.)
Also shown is the experimental energy (at the center of target) $E_\mathrm{c.m.}=233.2$~MeV
used in the GSI-TASCA experiment~\cite{khuyagbaatar2014}.
We note that the chosen experimental energy is $22.0$~MeV above the barrier $E_B$(tip)
and $8.6$~MeV above the barrier $E_B$(side).


\subsection{Energy dependence for central collision}

\begin{figure}[!htb]
\includegraphics*[width=8.6cm]{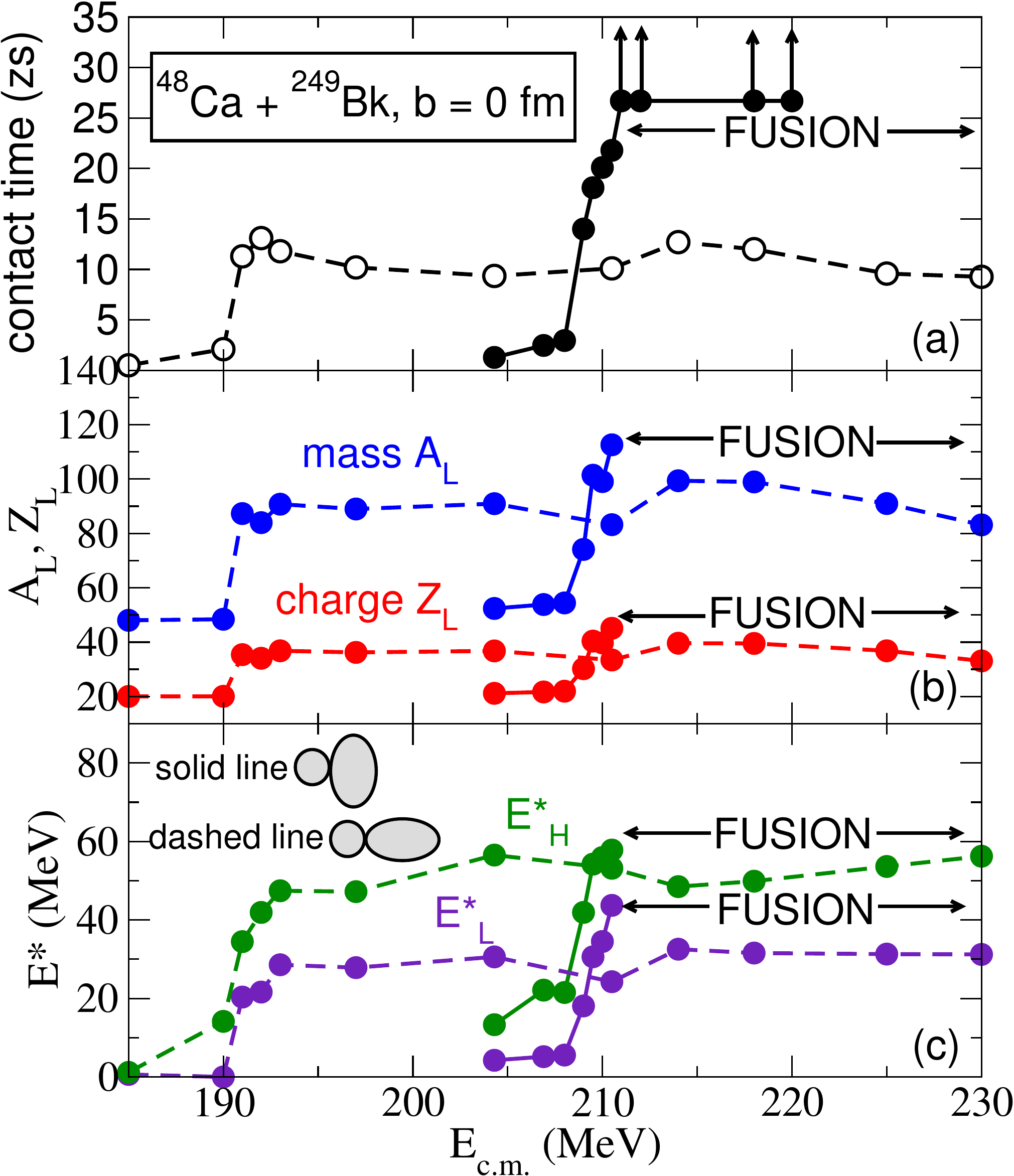}
\caption{\protect(Color online) (a) Contact time,
(b) mass and charge of the light fragment, and (c) excitation energy $E^{*}$
of the heavy and light fragments
as a function of $E_\mathrm{c.m.}$ for central collisions
of $^{48}$Ca with $^{249}$Bk. Solid lines are for the side
orientation of the deformed $^{249}$Bk nucleus, and dashed
lines are for the tip orientations.}
\label{fig:fig5}
\end{figure}

We define the contact time as the time interval between the time $t_1$
when the two nuclear surfaces 
(defined as isodensities with half the saturation density $\rho_0/2=0.08$~fm$^{-3}$) 
first merge into a single surface and
the time $t_2$ when the surface splits up again.  
Figure~\ref{fig:fig5} (a) shows the contact time as a function
of center-of-mass energy for central collisions of $^{48}$Ca with $^{249}$Bk calculated in TDHF.
For the tip orientation of the $^{249}$Bk nucleus (dashed line) we observe
contact times of order $10-12$~zs which are essentially constant over a wide range
of energies, $E_{\mathrm{c.m.}}=193-230$~MeV. Only at energies below the
potential barrier, $E_B$(tip)$=192.2$~MeV, do the contact times drop off
very rapidly because these events correspond to inelastic scattering
and few-nucleon transfer reactions. A dramatically different picture
emerges for the side orientation of the $^{249}$Bk nucleus (solid line):
At energies above the barrier $E_B$(side)$=205.4$~MeV, the contact times
rise very steeply with energy and reach values up to $22$~zs at
$E_{\mathrm{c.m.}}=210$~MeV. For energies above this value, TDHF predicts
fusion.
In our TDHF calculations, 
we have found fusion events at energies $E_{\mathrm{c.m.}}=211,212,218,220$~MeV for the side orientation.

Figure~\ref{fig:fig5} (b) shows the corresponding mass and charge of the light fragment.
We observe that the mass and charge transfer to the light fragment are
roughly proportional to the nuclear contact time. In particular,
for the side orientation of $^{249}$Bk, we find quasielastic collisions at energies below
$E_\mathrm{c.m.}=204$~MeV. Quasifission is limited to the narrow energy window
$E_\mathrm{c.m.}=209-211$~MeV, whereas for energies above $211$~MeV we find
fusion. Naturally, non-central impact
parameters can show quasifission in the range where we see fusion.
The quasifission results are very different for the tip orientation of $^{249}$Bk,
ranging over a much wider energy domain from $E_\mathrm{c.m.}=193$~MeV
to the highest energy of 230~MeV studied here,
with a lower maximum mass and charge transfer compared to
the side orientation of $^{249}$Bk.
Tip collisions clearly favor production of a heavy fragment near $^{208}$Pb (with a $^{91}$Rb light fragment) due to magic shell effects at all energies. 
A similar phenomenon was already observed in TDHF calculations of  reactions with $^{238}$U \cite{wakhle2014,sekizawa2016}.
In some cases, a light fragment with $N=50$ is  formed, indicating an influence of this magic number in the dynamics as well.

Recently, we have developed an extension to TDHF theory via the use
of a density constraint to calculate the excitation energy of {\it each fragment}
directly from the TDHF density evolution. This gives us new information on
the repartition of the excitation energy between the heavy and light fragments
which is not directly available in standard TDHF calculations 
unless one uses advanced projection techniques~\cite{sekizawa2015}.
In Fig.~\ref{fig:fig5} (c) we show the excitation energies of the two fragments
at internuclear distances of $26-30$~fm.
In collisions with the tip of the $^{249}$Bk nucleus, we find quasifission
excitation energies of $E^{*}_H=34-56$~MeV for the
heavy fragment and $E^{*}_L=20-32$~MeV for the light fragment, respectively.
For the side orientation, quasifission is only found in the narrow energy
window  $E_\mathrm{c.m.}=209-211$~MeV, with corresponding excitation energies of
$E^{*}_H=42-58$~MeV and $E^{*}_L=18-44$~MeV.

\begin{figure}[!htb]
\includegraphics*[width=8.6cm]{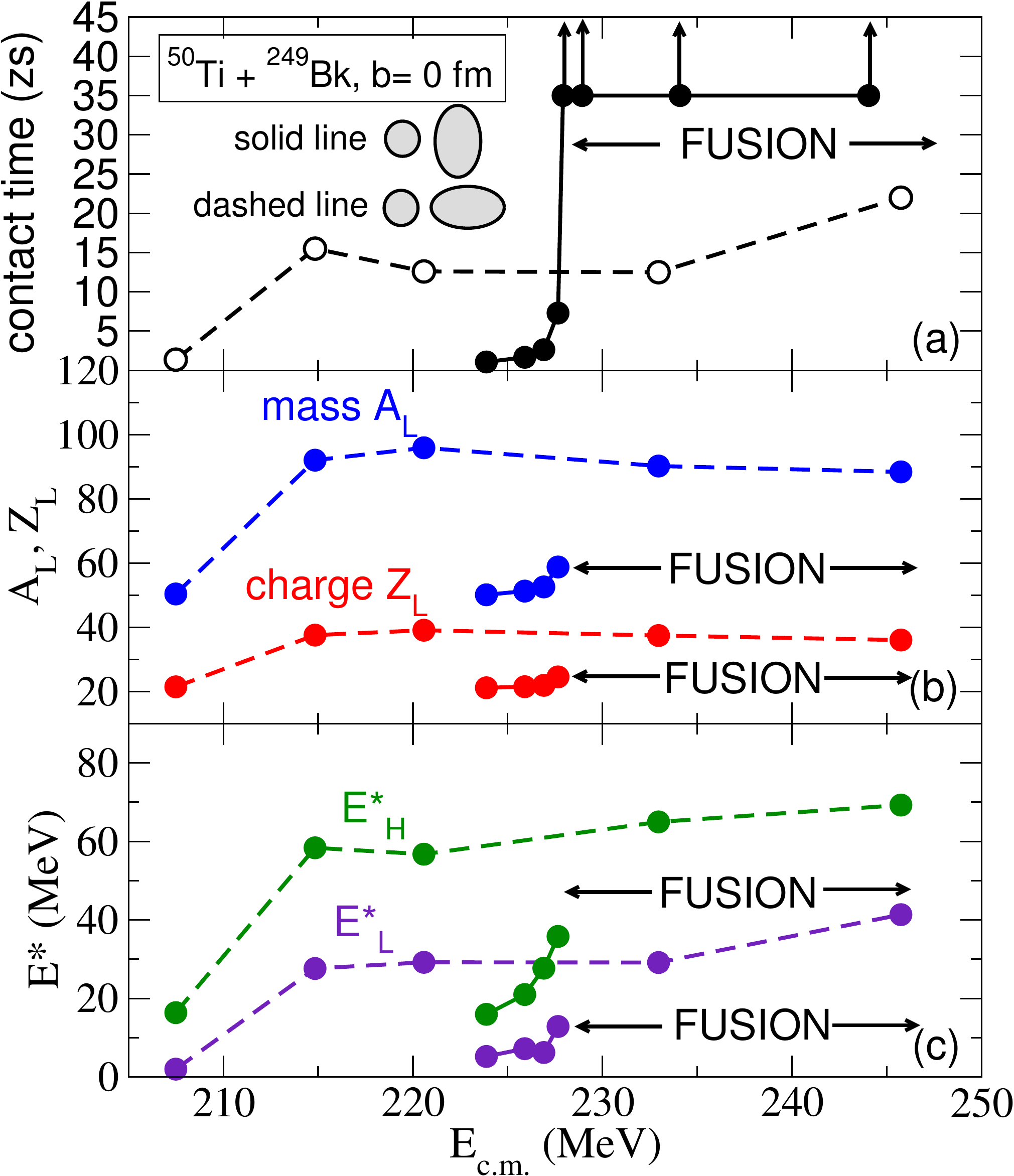}
\caption{\protect(Color online) (a) Contact time,
(b) mass and charge of the light fragment, and (c) excitation energy $E^{*}$
of the heavy and light fragments
as a function of $E_\mathrm{c.m.}$ for central collisions
of $^{50}$Ti with $^{249}$Bk. Solid lines are for the side
orientation of the deformed $^{249}$Bk nucleus, and dashed
lines are for the tip orientations.}
\label{fig:fig6}
\end{figure}

Figure~\ref{fig:fig6} shows the corresponding results for central collisions of
 $^{50}$Ti with $^{249}$Bk.
The contact times and the masses and charges of the light fragment show
a similar behavior as a function of the center-of-mass energy as compared
to the $^{48}$Ca $+^{249}$Bk reaction.
For the tip orientation, we find quasifission for
$E_\mathrm{c.m.}\ge214$~MeV, with
excitation energies of $E^{*}_H=57-69$~MeV for the
heavy fragment and $E^{*}_L=27-41$~MeV for the light fragment, respectively.
The mass and charge of the fragments indicate a strong influence 
of the shell effects in the $^{208}$Pb region, as in reactions with $^{48}$Ca. 
However, $N=50$ does not seem to play a role here. 
For the side orientation, we find inelastic and multi-nucleon transfer reactions
at energies $E_\mathrm{c.m.}=223-227$~MeV. Quasifission is confined to an
extremely narrow energy window around $E_\mathrm{c.m.}=227.4-227.7$~MeV, with
excitation energies of $E^{*}_H\simeq36$~MeV and $E^{*}_L\simeq13$~MeV. At energies
$E_\mathrm{c.m.}>228$~MeV, fusion sets in.


\subsection{Impact parameter dependence at fixed energy}

For the $^{48}$Ca+$^{249}$Bk system, we now examine the impact parameter
dependence of the same observables at a fixed energy of 
$E_{\mathrm{c.m.}}= 218$~MeV which is the highest energy
used in both the Dubna and GSI-TASCA experiments.
\begin{figure}[!htb]
\includegraphics*[width=8.6cm]{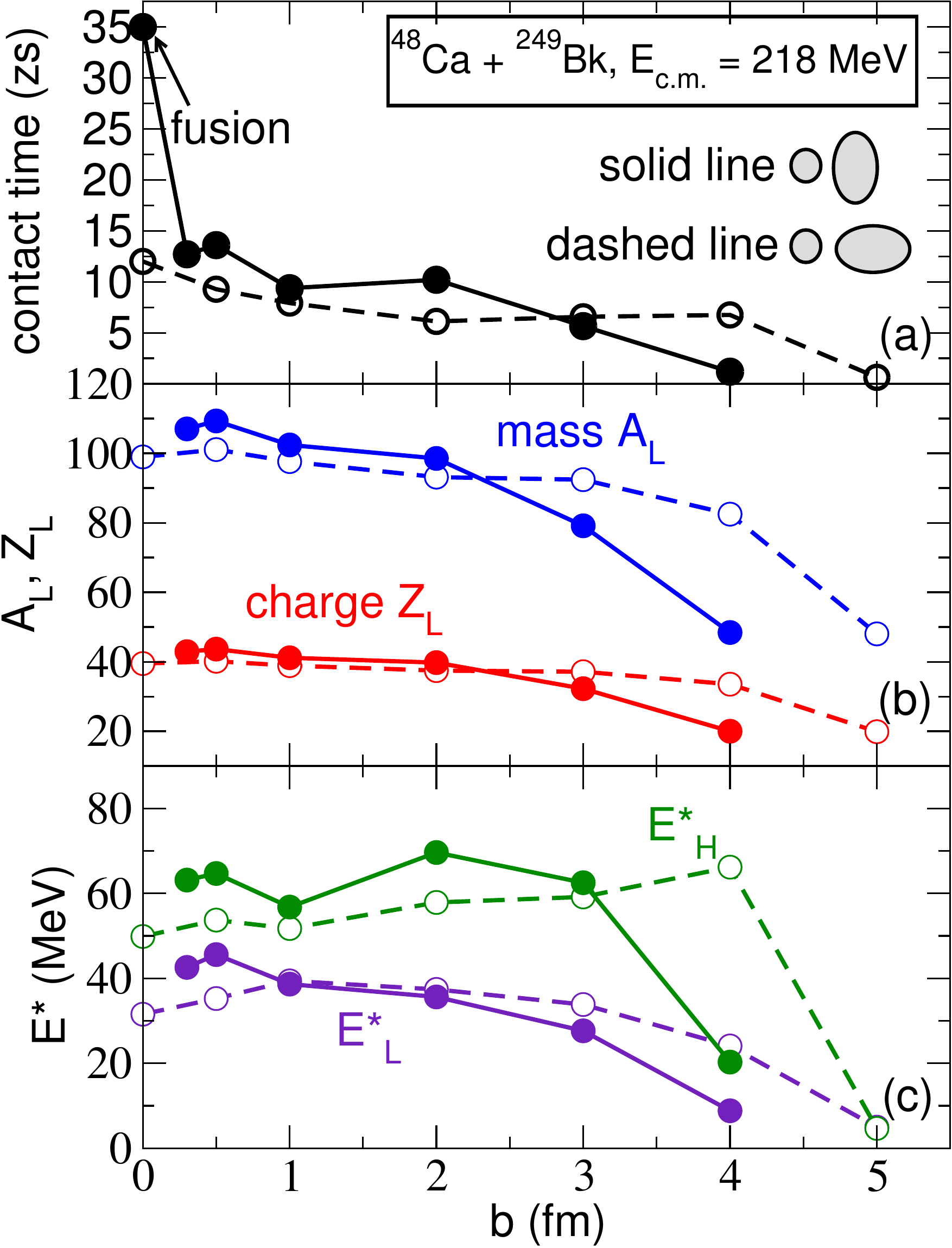}
\caption{\protect(Color online) (a) Contact time, (b) mass and charge of the light fragment,
and (c) excitation energy $E^{*}$ of the heavy and light fragments
for $^{48}$Ca+$^{249}$Bk as a function of impact parameter,
calculated at $E_{\mathrm{c.m.}}= 218$~MeV.}
\label{fig:fig7}
\end{figure}
Let us first consider collisions of $^{48}$Ca with the side of $^{249}$Bk.
Figure~\ref{fig:fig7} (solid lines) shows that fusion is only observed in the narrow impact parameter
region $b<0.3$~fm, as evidenced by contact times exceeding $35$~zs and 
a mononuclear shape without any neck formation. Quasifission reactions 
with contact times of $5.6-13.6$~zs are found
at impact parameters $b=0.3-3.0$~fm, with light fragment masses $A_L=80-109$ and
excitation energies $E^{*}_L=28-45$~MeV. Impact parameters $b>4$~fm yield deep-inelastic collisions (DIC),
multi-nucleon transfer and inelastic collisions.
\begin{figure}[!htb]
\includegraphics*[width=8.6cm]{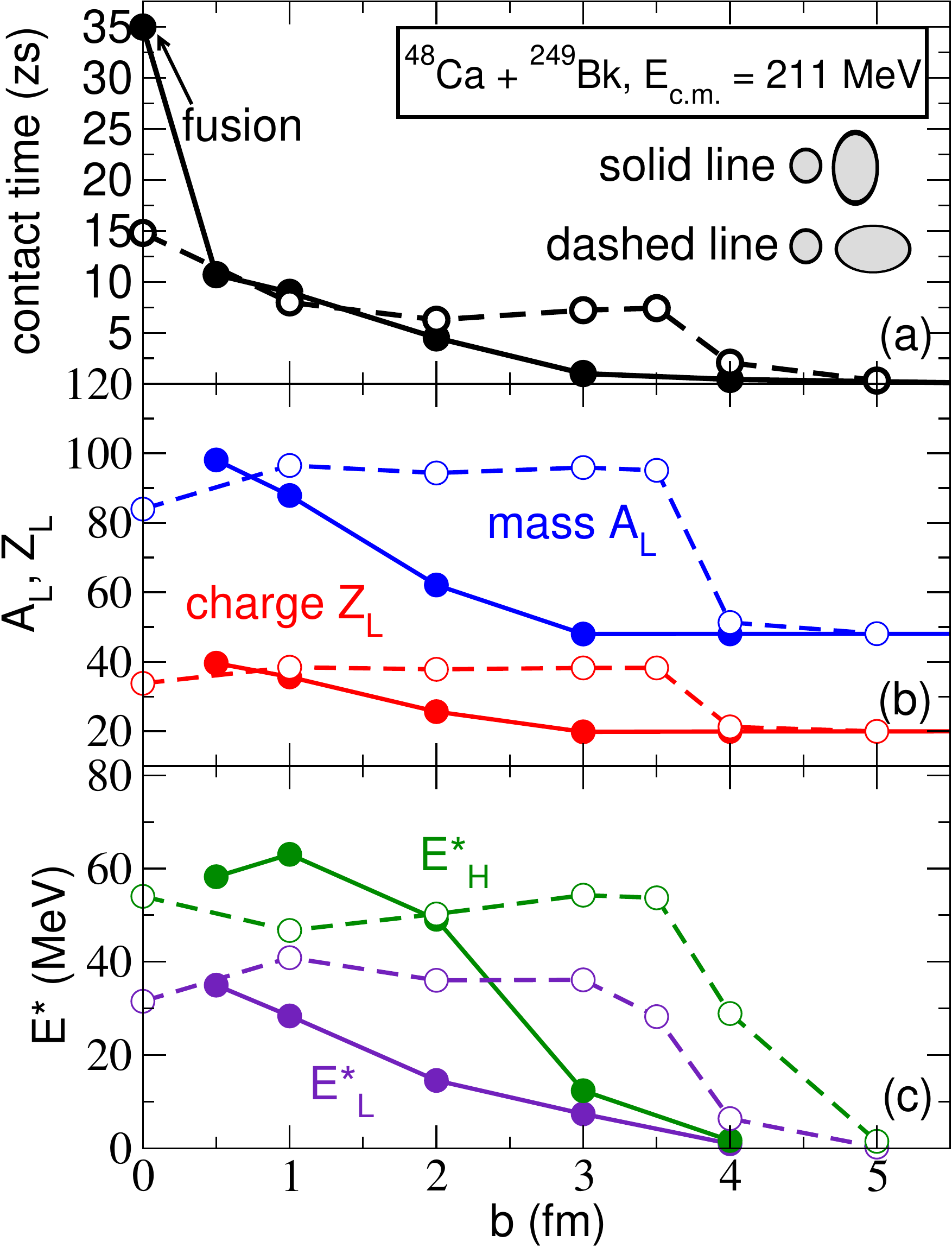}
\caption{\protect(Color online) Same as Fig.~\ref{fig:fig7}, but
calculated at a lower energy $E_{\mathrm{c.m.}}= 211$~MeV.}
\label{fig:fig8}
\end{figure}

It is interesting to note the atypical rise of the contact time between impact
parameters $b=1$~fm and $b=2$~fm, see Fig.~\ref{fig:fig7} (a). As shown in  
Fig.~\ref{fig:fig7} (b), for these impact parameters the light fragment is in the region of the neutron rich
$^{100}$Zr isotope. 
The microscopic evolution of the shell structure seems to have a tendency to
form a composite with a longer lifetime when the light fragment is in this region.
This was also discussed for the case of $^{40,48}$Ca$+^{238}$U quasifission study
of Ref.~\cite{oberacker2014}. In Ref.~\cite{oberacker2014} this was explained as
being due to the presence of strongly bound deformed isotopes of Zr in this
region~\cite{oberacker2003,blazkiewicz2005}.

Next we consider collisions of $^{48}$Ca with the tip of $^{249}$Bk (dashed lines in Fig.~\ref{fig:fig7}).
No fusion events are found for this initial orientation. 
Quasifission reactions with contact times of $6-12$~zs are found
at impact parameters $b=0-4$~fm, with light fragment masses $A_L=82-101$ and
excitation energies $E^{*}_L=24-39$~MeV. Impact parameters $b>5$~fm yield DIC,
multi-nucleon transfer and inelastic collisions. 

We have repeated these calculations at a lower center-of-mass energy of
$E_{\mathrm{c.m.}}= 211$~MeV. The results are shown in Figure~\ref{fig:fig8}.
For the tip orientation, all observables are quite similar to those obtained
at $E_{\mathrm{c.m.}}= 218$~MeV. However, for the side orientation, we find
that the contact time decreases more rapidly with impact parameter than at
higher energy. As a result, both mass transfer and fragment excitation
energies also decrease faster. Fusion is found for impact parameters
 $b<0.5$~fm (for the side orientation of $^{249}$Bk only).

\begin{figure}[!htb]
\includegraphics*[width=8.6cm]{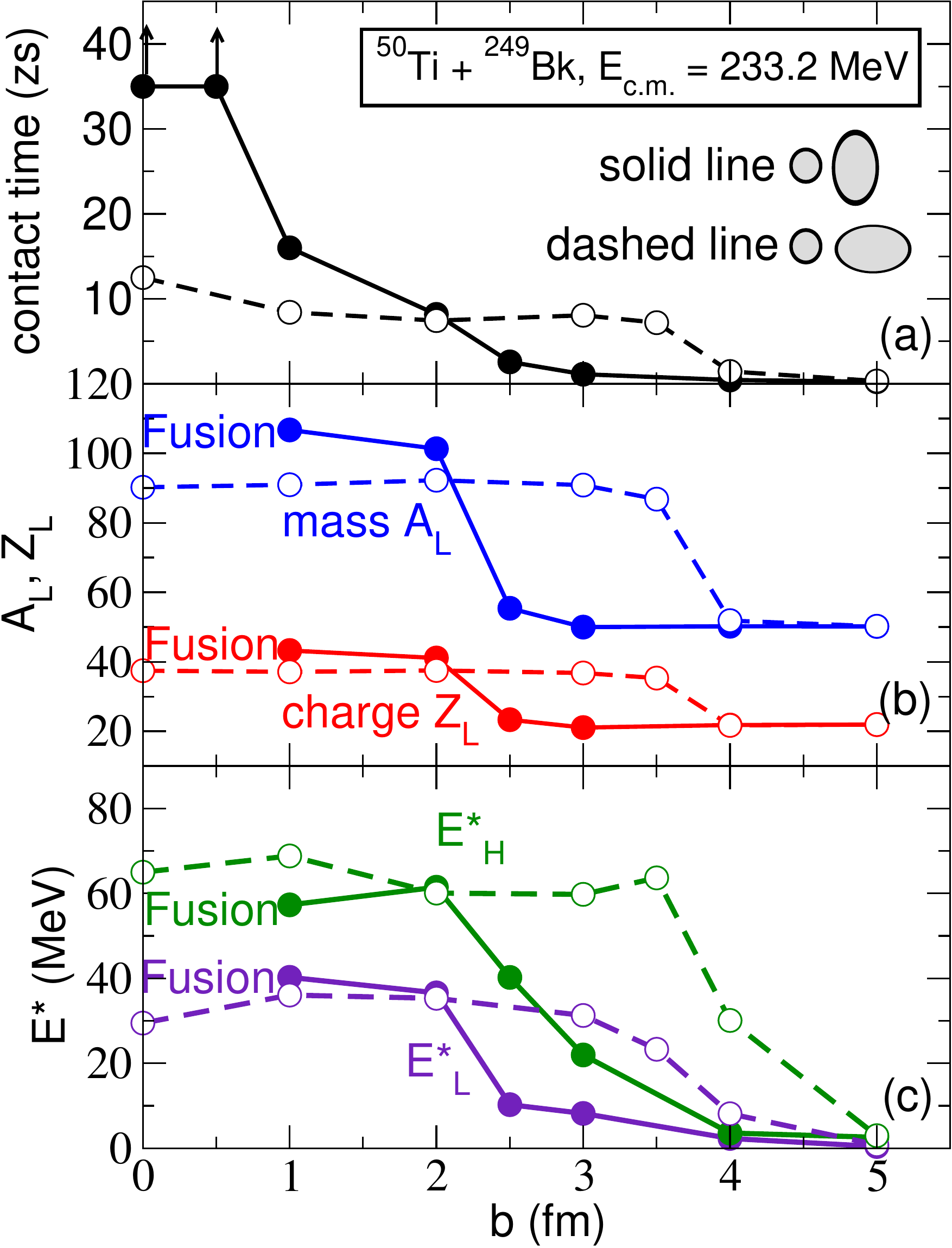}
\caption{\protect(Color online) (a) Contact time, (b) mass and charge of the light fragment,
and (c) excitation energy $E^{*}$ of the heavy and light fragments
for $^{50}$Ti+$^{249}$Bk as a function of impact parameter,
calculated at $E_{\mathrm{c.m.}}= 233.2$~MeV.
}
\label{fig:fig9}
\end{figure}

Figure~\ref{fig:fig9} shows results for the $^{50}$Ti+$^{249}$Bk system at
$E_{\mathrm{c.m.}}= 233.2$~MeV as a function of impact parameter.
Let us first consider collisions with the side of $^{249}$Bk (solid lines).
Fusion is observed at impact parameters $b=0$~fm and  $b=0.5$~fm. 
Quasifission reactions with contact times of $8-16$~zs are found
at impact parameters $b=1$~fm and at $b=2$~fm, with light fragment masses $A_L=101-107$ and
excitation energies $E^{*}_L=36-40$~MeV. Impact parameters $b>2.5$~fm yield
DIC, multi-nucleon transfer and inelastic collisions.

Now we consider collisions with the tip of $^{249}$Bk (dashed lines).
No fusion events are found for this initial orientation. 
Quasifission reactions with contact times of $7.4-12.5$~zs are found
at impact parameters $b=0-3$~fm, with light fragment masses $A_L=90-92$ and
excitation energies $E^{*}_L=29-36$~MeV. Impact parameters $b>3.5$~fm produce
DIC, multi-nucleon transfer, and inelastic collisions. 

\begin{figure}[!htb]
\includegraphics*[width=8.6cm]{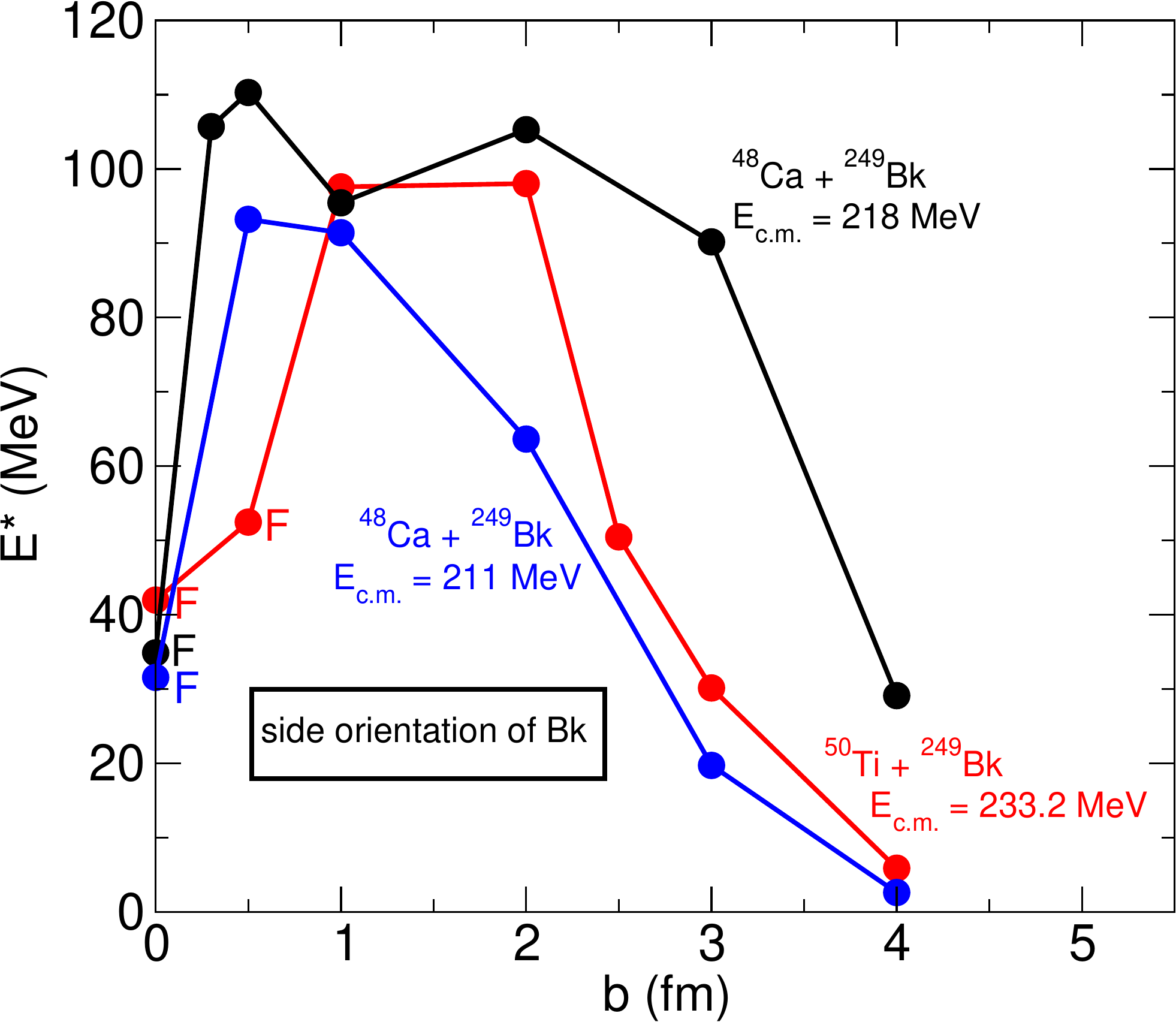}
\caption{\protect(Color online) Comparison of the total excitation energy
for the systems $^{48}$Ca+$^{249}$Bk and $^{50}$Ti+$^{249}$Bk (side orientation)
as a function of impact parameter. The points corresponding to fusion events
are indicated by a letter F.}
\label{fig:fig10}
\end{figure}

Experiments at Dubna and at GSI-TASCA have produced several isotopes of 
superheavy element $117$ with cross-sections of $2-3$ picobarns in the reaction
$^{48}$Ca +$^{249}$Bk. However, attempts to synthesize isotopes of element $119$ 
in the reaction $^{50}$Ti+$^{249}$Bk have been unsuccessful so far.
One possible reason could be different excitation energies in these systems.
In order to investigate this conjecture, we have calculated the total
excitation energy for both systems as a function of impact parameter. This
quantity can be calculated with the DC-TDHF method for both fusion and quasifission.
The results are displayed in Figure~\ref{fig:fig10} for the side orientation of 
$^{249}$Bk. 
The total excitation energy is shown as a function of impact parameter for the
$^{48}$Ca +$^{249}$Bk system at two center-of-mass energies, $E_{\mathrm{c.m.}}= 211$~MeV
and $E_{\mathrm{c.m.}}= 218$~MeV. Naturally, the excitation energy increases with
increasing $E_{\mathrm{c.m.}}$.
Also shown is the total excitation energy of
$^{50}$Ti+$^{249}$Bk, calculated at the GSI-TASCA energy of $E_{\mathrm{c.m.}}= 233.2$~MeV.
The most important region is the region of small impact parameters
where fusion occurs (for the side orientation of $^{249}$Bk only).
We find the interesting result that the total excitation energy of both systems is almost
identical at impact parameters $b=0$~fm (fusion) and $b=1$~fm (QF). 
For impact parameters $b>1.5$~fm, the total excitation energy of the 
$^{50}$Ti+$^{249}$Bk system is found to be
in between the two curves calculated for $^{48}$Ca +$^{249}$Bk.
We conclude that the excitation energy of the fused system or of the quasifission fragments does not exhibit strong differences between $^{48}$Ca and $^{50}$Ti induced reactions.


\subsection{Mass-angle distributions}

In this section we study mass-angle distributions (MADs)
arising from quasifission. 
MADs have proven to be an efficient experimental tool to understand quasifission dynamics and how this mechanism is affected by the structure of the reactants \cite{toke1985,shen1987,rafiei2008,thomas2008,hinde2008,rietz2011,lin2012,simenel2012b,yadav2012,rietz2013,williams2013,wakhle2014,hammerton2015,prasad2015,prasad2016,shamlath2016}. 
TDHF calculations can help the analysis and interpretation of experimental MADs \cite{simenel2012b,wakhle2014,hammerton2015,prasad2016}.

\begin{figure}[!htb]
\includegraphics*[width=8.6cm]{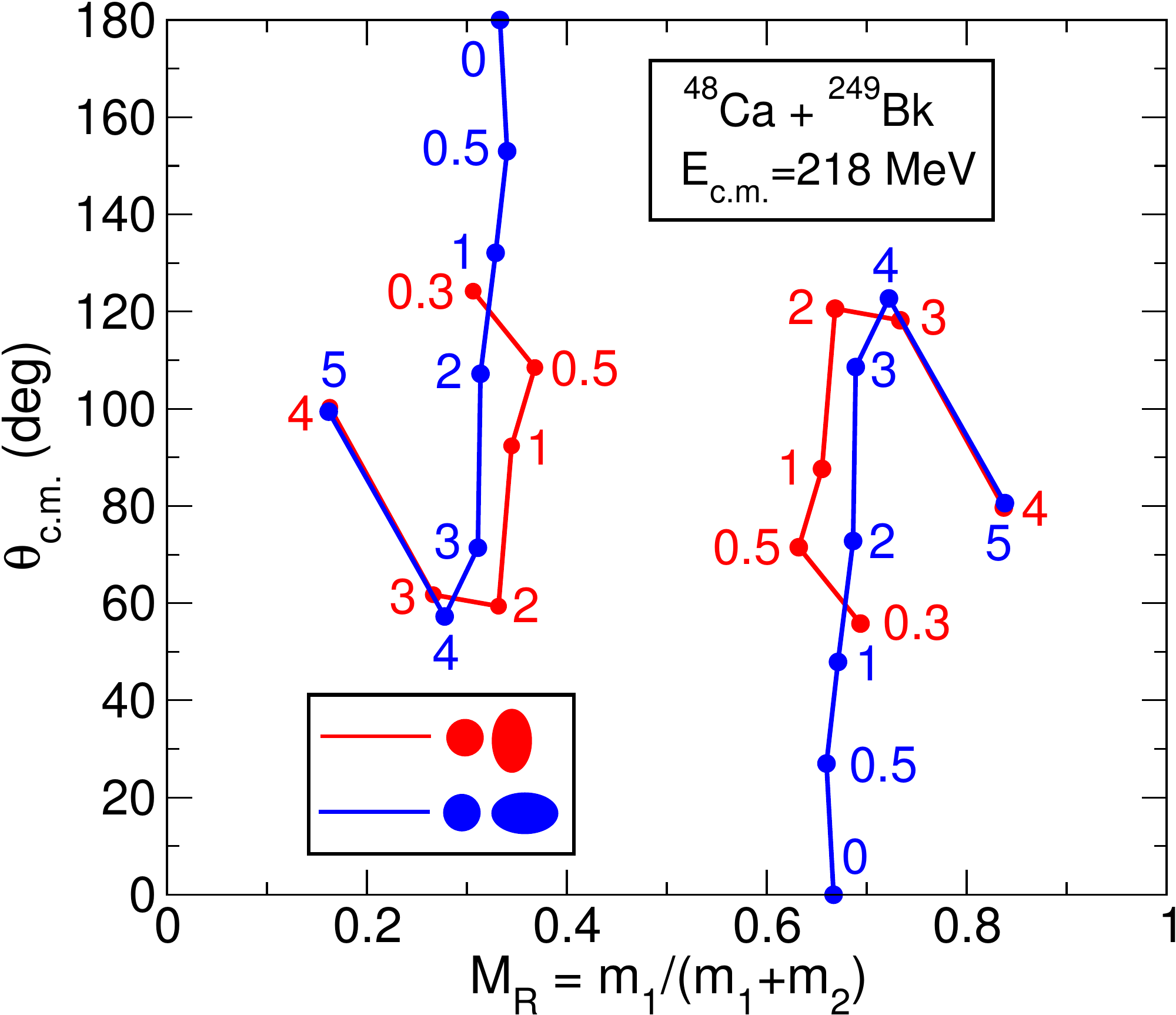}
\caption{\protect(Color online) Mass-angle distribution for the $^{48}$Ca+$^{249}$Bk system
at $E_\mathrm{c.m.}=218$~MeV, calculated for tip and side orientations of the $^{249}$Bk nucleus.
The impact parameters (in units of fm) are indicated by the numbers next to each
data point.}
\label{fig:fig11}
\end{figure}

The MAD is obtained by plotting the
scattering angle $\theta_{c.m.}$ as a function of the mass ratio $M_R=m_1/(m_1+m_2)$
where $m_1$ and $m_2$ are the masses of the fission-like fragments.
In Fig.~\ref{fig:fig11} we show TDHF calculations of mass-angle distributions for
$^{48}$Ca+$^{249}$Bk at $E_\mathrm{c.m.}=218$~MeV.
The MAD regions near $M_R=0.16$ and $M_R=0.84$ correspond to
quasielastic and deep-inelastic reactions. In the region $0.25 < M_R <0.75$
fissionlike fragments are observed.

For the quasifission events which occur at time scales between $5.6$~zs  and $13.6$~zs,
our TDHF calculations show a strong correlation between scattering angle and mass
ratio. The reason for this correlation is that the mass
transfer between the two fragments increases with the rotation (contact) time
(see Figure~\ref{fig:fig7}a,b) which in turn impacts
 the scattering angle. 
 Hence, the MADs for quasifission
events can be used as a clock for the rotation period of the system \cite{toke1985,rietz2011}.

For the tip orientation of the $^{249}$Bk
nucleus (blue curve in Fig.~\ref{fig:fig11}) TDHF shows quasifission at impact parameters
$b=0-4$~fm and a deep-inelastic reaction at $b=5$~fm. No fusion events are
predicted by TDHF for the tip orientation. On the other hand, for the side
orientation of the $^{249}$Bk nucleus (red curve) the TDHF calculations
show fusion at impact parameter $b=0$~fm, quasifission at impact parameters
$b=0.3-3.0$~fm, and a deep-inelastic reaction at $b=4$~fm. In general, collisions
with the side of $^{249}$Bk yield an increase in the mass ratio for quasifission. The maximum
value for the light fragment, $M_R=0.368$, is obtained at impact parameter $b=0.5$~fm.
Note that as a result of the single-Slater-determinant approximation, TDHF is a
deterministic theory that will provide us only with the most probable reaction products
for the MADs rather than with the full mass distribution.

In fusion-fission reactions a compound nucleus is formed which subsequently
decays by  fission
at a time-scale that is much longer than observed in quasifission, with no memory of the entrance channel and therefore
no mass-angle correlation. In experiments,   fission fragments are usually more symmetric than in quasifission, producing a peak around $M_R=0.5$. Even though our TDHF
calculations predict fusion for the small impact parameter range
$b<0.3$~fm (and only for the
side orientation of the $^{249}$Bk nucleus), it is not possible to obtain a fully equilibrated nucleus undergoing fission in TDHF calculations because of limitations of the mean-field approach.

\begin{figure}[!htb]
\includegraphics*[width=8.6cm]{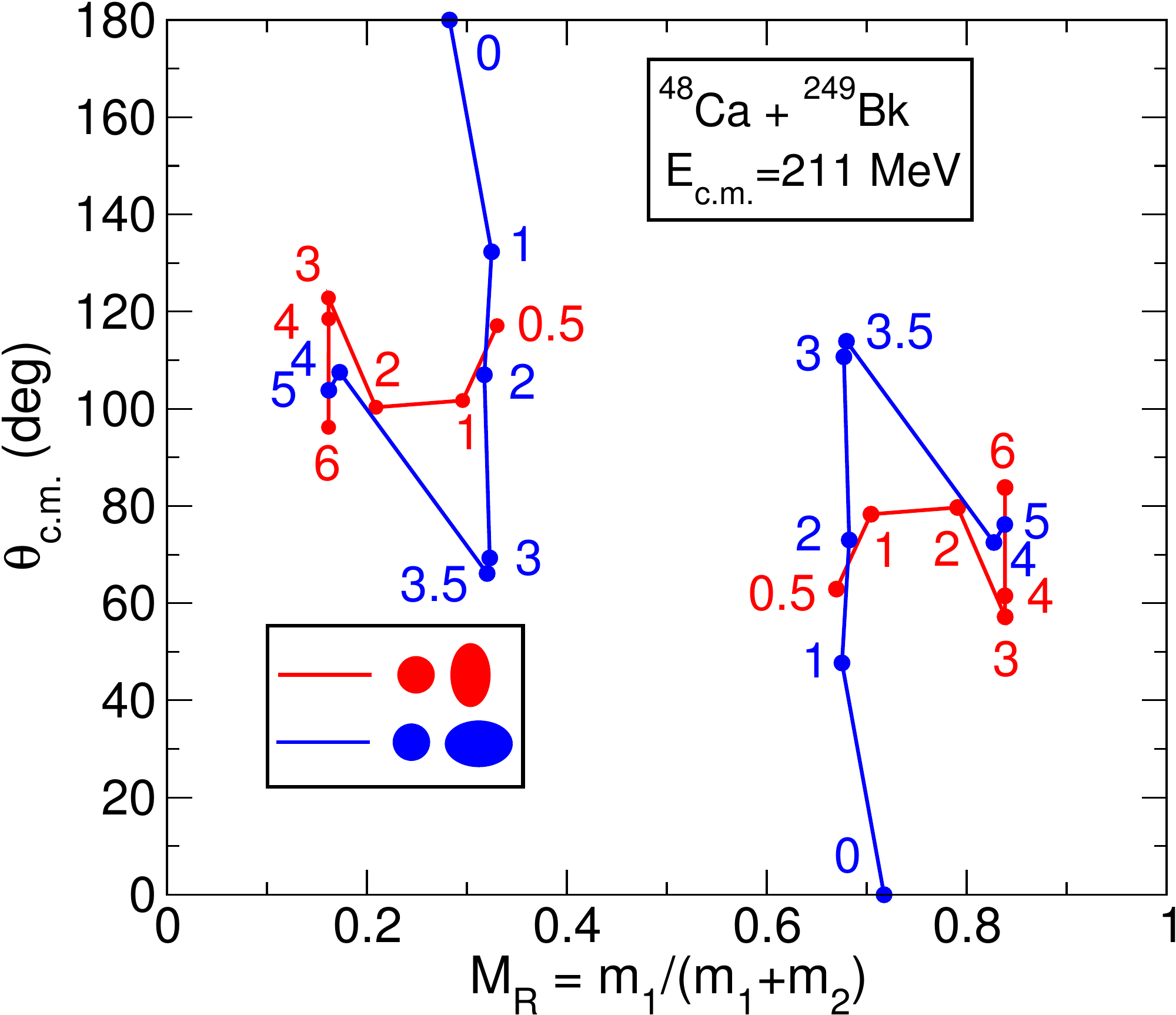}
\caption{\protect(Color online) Same as Fig.~\ref{fig:fig11}, but at a lower energy
$E_\mathrm{c.m.}=211$~MeV.}
\label{fig:fig12}
\end{figure}

In Fig.~\ref{fig:fig12} we show TDHF calculations of the mass-angle distribution
for the same system, but at a lower energy $E_\mathrm{c.m.}=211$~MeV.
The MAD for the tip orientation of the $^{249}$Bk nucleus (blue curve) looks quite
similar to the one obtained at higher energy. However, for the side orientation (red curve)
we find a  different mass-angle distribution: the scattering angles for the light fragment
are confined to a small region $\theta_{c.m.} = 96-123$~deg, and the fragments are more asymmetric than at $E_\mathrm{c.m.}=218$~MeV.

\begin{figure}[!htb]
\includegraphics*[width=8.6cm]{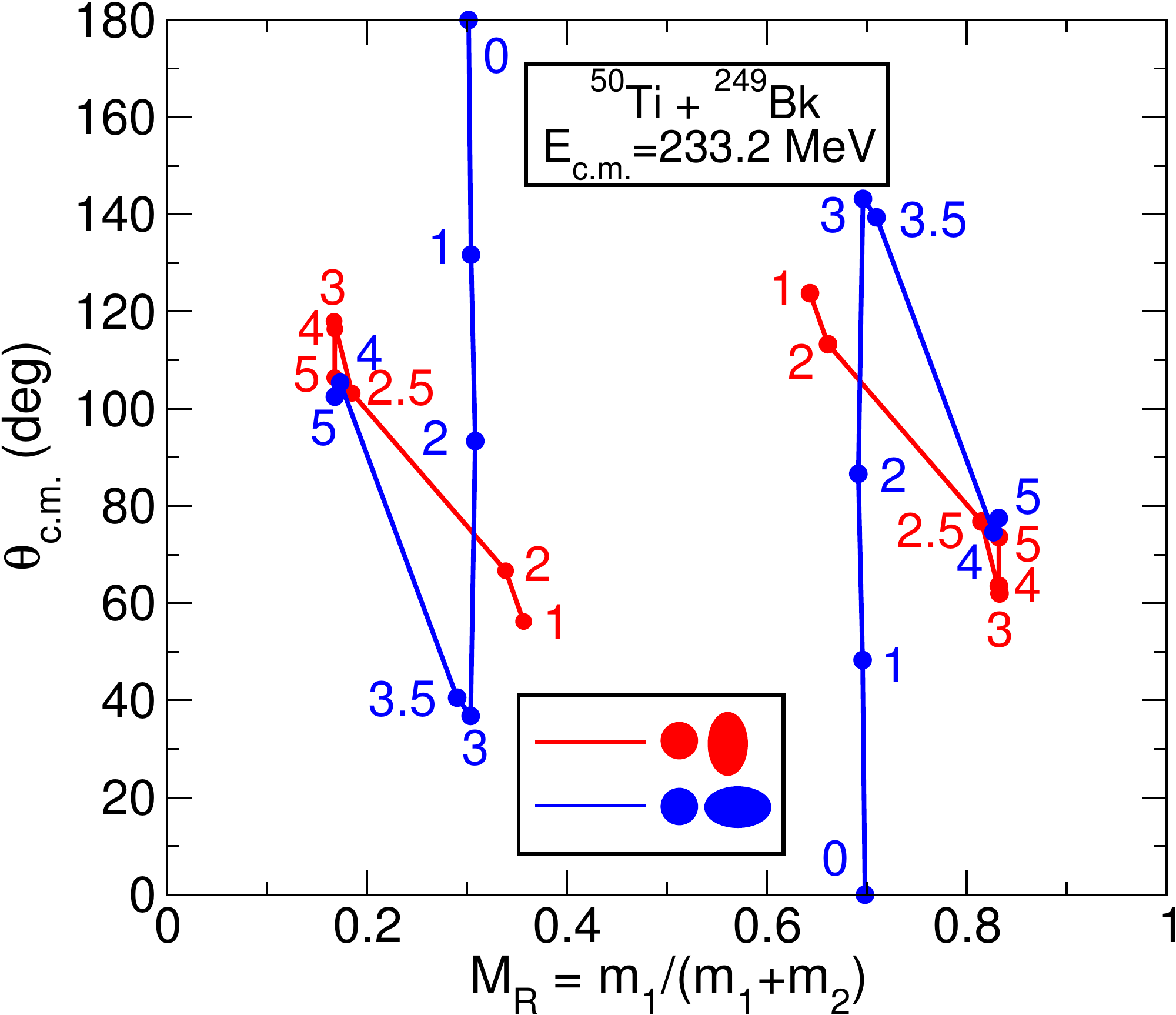}
\caption{\protect(Color online) Mass-angle distribution for $^{50}$Ti+$^{249}$Bk
at $E_\mathrm{c.m.}=233$~MeV, calculated for tip and side orientations of the $^{249}$Bk nucleus.
The impact parameters (in units of fm) are indicated by the numbers next to each
data point.}
\label{fig:fig13}
\end{figure}

In Fig.~\ref{fig:fig13} we show TDHF calculations of quasifission
mass-angle distributions for
$^{50}$Ti+$^{249}$Bk at $E_\mathrm{c.m.}=233$~MeV, corresponding to the two
orientations of the $^{249}$Bk nucleus.
The MAD for the tip orientation (blue curve) are very similar to the result
obtained for $^{48}$Ca+$^{249}$Bk at $E_\mathrm{c.m.}=218$~MeV (see Fig.~\ref{fig:fig11}).
The MAD  for the side orientation (red curve) also shows similarities with the $^{48}$Ca induced one at 218~MeV, 
with fragments  produced  in a similar angular range ($60-120$~deg.) 
and with a maximum $M_R$ for the light fragment extending almost to 0.4. 
In the details, however, differences are observed on the position in the MADs of events associated with specific impact parameters.


\subsection{Mass-TKE distributions}

Correlations between mass and total kinetic energy (TKE) of the fragments have often been measured in experimental studies of quasifission \cite{toke1985,shen1987,hinde1992,itkis2004,itkis2007,knyazheva2007,prokhorova2008,kozulin2010,itkis2011,kozulin2014}.
Plots of fragment mass versus TKE are often used to separate quasi-elastic events to fully damped events such as quasifission and fusion-fission. 
In between, deep-inelastic collisions are characterised by a partial damping of the initial kinetic energy and a relatively small (compared to quasifission) mass transfer. 
Fully damped events are expected to have a TKE close to the Viola systematics for fission fragments \cite{viola1985,hinde1987}.

In TDHF, the TKE of the fragments is simply obtained from the exit channel of the collision. 
For well separated fragments, it is straightforward to compute 
the kinetic energy of each fragment ($i=1,2$) at time $t$ according to
$$T_i(t)=\frac{1}{2}M_i\left(\frac{dR_i(t)}{dt}\right)^2,$$ 
where $M_i$ is the final mass of the fragment $i$ (neglecting nucleon emission) and $R_i$ its distance from the center of mass of the total system.
Although the fragments do not interact  anymore via the strong nuclear interaction, they are close enough feel the Coulomb repulsion from the other fragment. 
The TKE is then estimated by the sum of the kinetic energy of the fragments after their separation and their Coulomb potential energy assuming that the fragments are point like charges,
$$
TKE\simeq T_1(t)+T_2(t)+\frac{Z_1Z_2e^2}{R(t)},
$$
where $R(t)=R_1(t)+R_2(t)$.
\begin{figure}[!htb]
\includegraphics*[width=8.6cm]{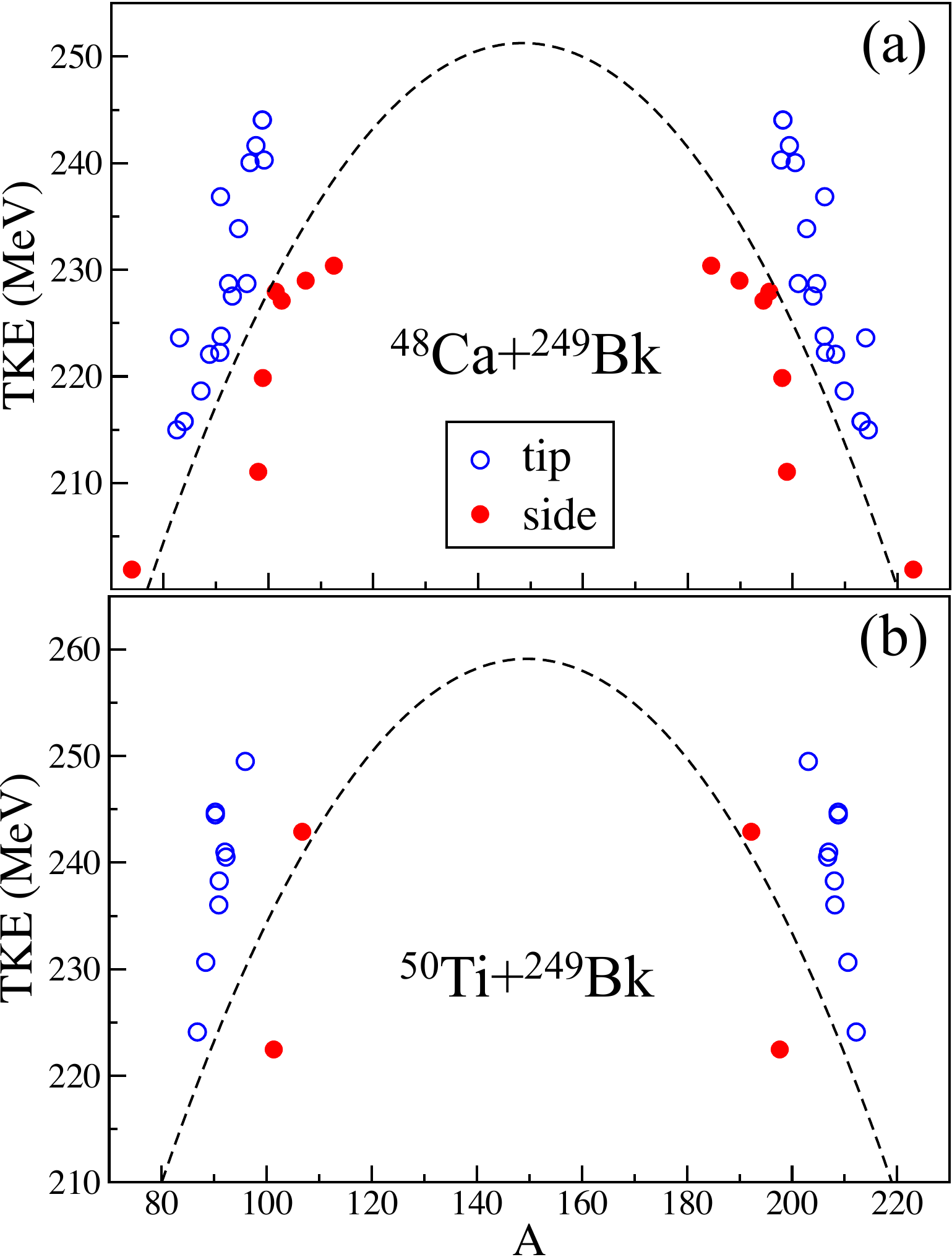}
\caption{\protect(Color online) TKE-mass correlations in (a) $^{48}$Ca+$^{249}$Bk and (b) $^{50}$Ti+$^{249}$Bk.
Tip (side) orientations are plotted with open (filled) symbols. The dashed line is the Viola systematics assuming that the fragments have the neutron-to-proton ratio of the compound nucleus.}
\label{fig:fig14}
\end{figure}

Figures~\ref{fig:fig14}(a) and (b) show a compilation of the mass-TKE distributions obtained in $^{48}$Ca,$^{50}$Ti+$^{249}$Bk TDHF calculations, respectively.
The figures also show the TKE expected from the Viola systematics accounting for fragment mass asymmetry \cite{hinde1987} and assuming that the fragments have the same $N/Z$ ratio as the compound nucleus (dashed lines).
Overall, we observe that the TKE are distributed around the Viola estimates, indicating that  most of the relative kinetic energy has been dissipated in the collision. 
However, the distributions associated with side and tip orientations are well separated, with the side (tip) collisions leading  essentially to a TKE below (above) the Viola systematics.

\begin{figure}[!htb]
\includegraphics*[width=8.6cm]{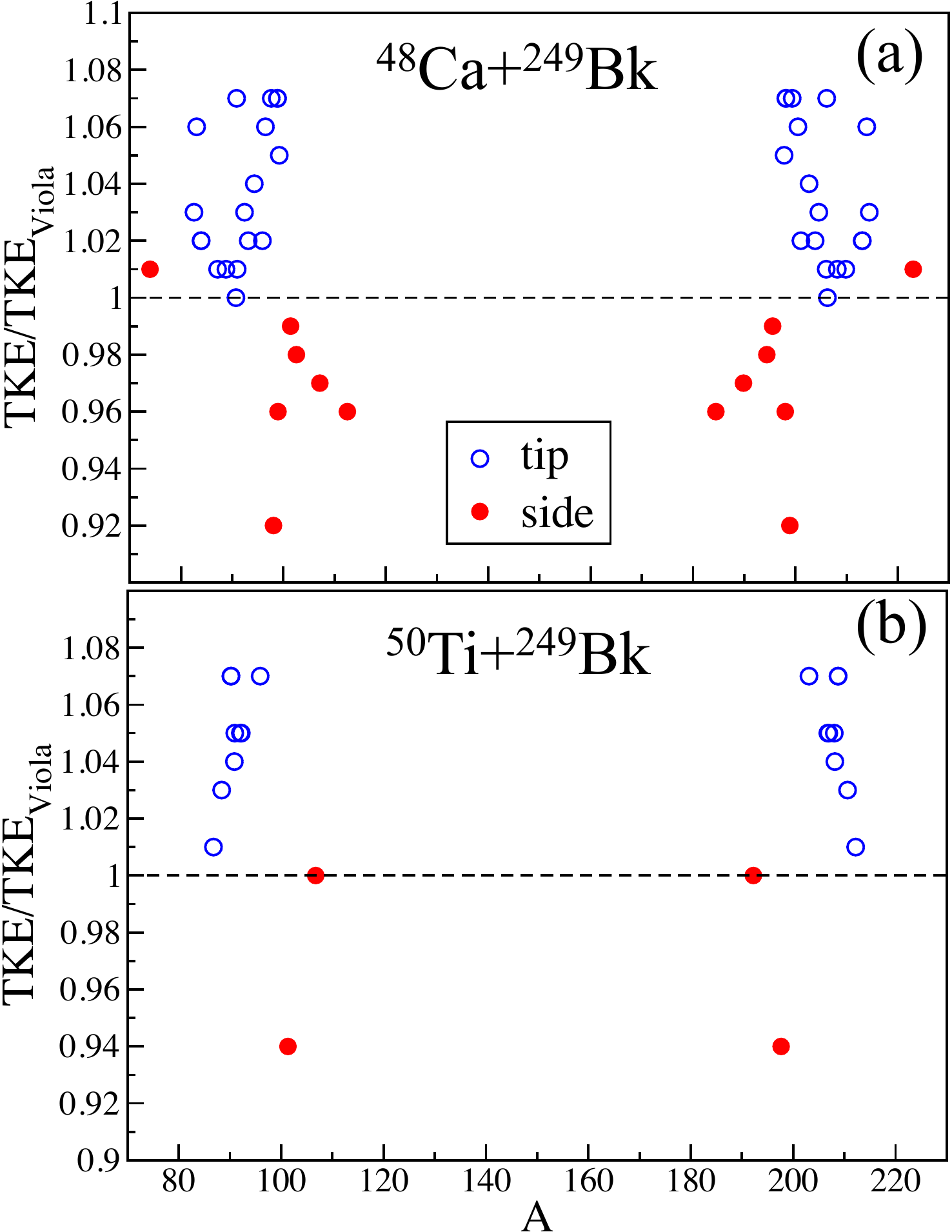}
\caption{\protect(Color online) Same as Fig.~\ref{fig:fig14} but with the TKE normalized to the TKE from Viola systematics using the masses and charges of the fragments in the exit channel.}
\label{fig:fig15}
\end{figure}

One could argue that assuming that the fragments have the same $N/Z$ as the compound nucleus is a crude approximation, in particular for systems with large asymmetry in the exit channels as observed here.
Therefore, we have also computed the TKE according to the Viola estimate using the masses and charges of the fragments in the exit channel obtained from TDHF.
The results are plotted in Figs.~\ref{fig:fig15}(a) and (b)  which show the ratio of the TDHF final TKE over the TKE from Viola estimate with TDHF mass and charge partitions. 
The previous conclusion are still valid, i.e., the tip (side) orientations are associated with more (less) final TKE than the Viola systematics.
This means that less damping occurs in collisions with the tip than with the side. 
This conclusion, however, does not depend on if the projectile is a $^{48}$Ca or a $^{50}$Ti, indicating again that the reaction dynamics is relatively similar in both systems. 


\section{Conclusions}\label{sec:conclusions}
The Time-Dependent Hartree-Fock (TDHF) theory provides a dynamic quantum
many-body description of nuclear reactions. The only input is the effective nucleon-nucleon
interaction (Skyrme) which is fitted to the static properties of a few nuclei,
otherwise there are no adjustable parameters. TDHF has proven to be a valuable tool for
elucidating some of the underlying physics of heavy-ion reactions in the vicinity of the
Coulomb barrier. In this paper, we have studied the transition between various reaction mechanisms including fusion, quasifission, deep-inelastic
collisions, and
quasi-elastic reactions in collisions of $^{48}$Ca+$^{249}$Bk
and $^{50}$Ti+$^{249}$Bk which have been used to synthesize elements
$Z=117,119$. Quasifission is the primary reaction mechanism that limits the
formation of superheavy nuclei.

In addition, heavy-ion interaction potentials are obtained
with the Density-Constrained Time-Dependent Hartree-Fock (DC-TDHF) method.
Because of the prolate deformation of the Bk nucleus, these potentials
(and other observables) depend strongly on the relative orientation of $^{249}$Bk.
In particular, we present results for the ``tip'' and ``side'' orientation.
Using TDHF, we calculate nuclear contact times, masses and charges of the two
fragments, and their pre-compound excitation energies. Specifically, we study
the energy-dependence of these quantities for central collisions, and we calculate the impact
parameter dependence at selected fixed energies. Finally, we present results for mass-angle
and mass-TKE distributions. 
The orientation of the actinide plays a crucial role on these observables. 

In agreement with experiments at Dubna and at GSI-TASCA, our TDHF and DC-TDHF
calculations predict fusion in the reactions $^{48}$Ca +$^{249}$Bk resulting
in isotopes of element $117$. While experimental attempts at GSI-TASCA to synthesize
element $119$ in the reaction $^{50}$Ti+$^{249}$Bk have so far not been successful,
the TDHF calculations do find fusion in this system also. 
In fact, our calculations do not show significantly different behaviors of the entrance channel dynamics between the two projectiles.


\begin{acknowledgments}
We thank Y. Oganessian, W. Loveland, K. P. Rykaczewski, and D. J. Hinde for stimulating discussions.
This work has been supported by the U.S. Department of Energy under grant No.
DE-SC0013847 with Vanderbilt University and by the
Australian Research Councils Future Fellowship (project number FT120100760) and Discovery Projects (project number DP160101254) funding schemes.
\end{acknowledgments}


\bibliography{VU_bibtex_master.bib}


\end{document}